\documentclass
[preprint,showpacs,byrevtex,prb,10pt,letterpaper,twocolumn]{revtex4}%
\usepackage{amsfonts}
\usepackage{amsmath}
\usepackage{amssymb}
\usepackage[dvips,final]{graphicx}%
\begin{document}
\preprint{ }
\title{Microwave-Enhanced hopping-conductivity; a non-Ohmic Effect}
\author{Z. Ovadyahu}
\affiliation{Racah Institute of Physics, The Hebrew University, Jerusalem 91904, Israel }
\pacs{72.80.Ng 73.50.Fq 72.20.Ee}

\begin{abstract}
Hopping conductivity is enhanced when exposed to microwave fields (Phys. Rev.
Lett., \textbf{102}, 206601, 2009). Data taken on a variety of
Anderson-localized systems are presented to illustrate the generality of the
phenomenon. Specific features of these results lead us to conjecture that the
effect is due to a field-enhanced hopping, which is the high frequency version
of the non-Ohmic effect, well known in the dc transport regime. Experimental
evidence in support of this scenario is presented and discussed. It is pointed
out that existing models for non-Ohmic behavior in the hopping regime may, at
best, offer a qualitative explanation to experiments. In particular, they
cannot account for the extremely low values of the threshold fields that mark
the onset of non-Ohmic behavior.

\end{abstract}
\maketitle

\section{Introduction}

Non-Ohmic effects are commonly encountered in hopping conductivity. Actually,
it is difficult to maintain linear-response conditions in these systems. The
degree of non-Ohmicity could be tamed by reducing the potential drop across
the sample but the need to ensure good signal-to-noise makes this harder as
the temperature gets lower. Deviations from linear response may be expected
when the applied field $F_{c}$ obeys:%

\begin{equation}
F_{c}\gtrsim\frac{k_{B}T}{eL} \label{1}%
\end{equation}

Here, $k_{_{B}}$ is Boltzmann constant; e is the electron charge, and $T$ the
temperature. $L$ is the spatial scale over which the electron gains energy
from the field $F$ before it is dissipated into the bath, usually by phonon
emission, and therefore $L$ is a function of temperature, typically of the
form $_{L}\propto T^{-p}$ ( $p$ is a number between 1to 4 depending on the
material, temperature range, and dimensionality). Modifications to the
conductance by a sufficiently large dc field were studied by several, somewhat
different theoretical approaches \cite{1,2,3,4}. These models however
predicted a similar result; the field induces an excess conductance $\Delta G$
that, at intermediate field-strength, is exponential with ($eFL/k_{_{B}}%
T$)$^{\gamma}$ with $\gamma$=1 or 2. Intermediate fields are fields in the
range: $F_{c}$%
$<$%
$F$%
$<$%
$F_{h}$. The high-field limit $F_{h}$ (where the conductance becomes
independent of temperature \cite{4,5}), is defined by $F_{h}\gtrsim\frac
{k_{B}T}{e\xi}$, where $\xi$ is the localization length.

Much less attention has been given to the effect of a long-wavelength
electromagnetic radiation on hopping conductivity. Of particular interest is
where the wavelength is larger than the sample-size \cite{6} which can be
readily implemented by using microwaves (MW) radiation. Ben-Chorin et al
observed that exposing the sample to a MW source enhanced its hopping
conductance and, perhaps naturally, associated the effect with heating
\cite{7}. However, systematic studies, exploiting some unique properties of
electron-glasses \cite{8}, suggested that this is a non-equilibrium effect. A
persistent feature of the results, which made it difficult to reconcile the
effect with heating, is a sub-linear dependence of $\Delta G$ on the
microwaves power $P$.

In this paper we report on results obtained by similar measurements on films
of five different hopping systems; granular-aluminum, In$_{\text{2}}%
$O$_{\text{3-x}}$ (crystalline indium-oxide), In$_{\text{x}}$O (amorphous
indium-oxide), beryllium, and GaAs. In all cases $\Delta G$ at $T\approx$4K
turns out to be sub-linear with the MW power in agreement with the results of
previous studies. It is further shown that the functional dependence of
$\Delta G(P)$ is consistent with the current-voltage characteristics measured
independently (at low frequency) on the same samples. The MW-enhanced
conductance is then conjectured to be a non-Ohmic effect, of a similar nature
as the field assisted-hopping phenomenon. This is a generic mechanism and
should be obeyed by any system once driven away from its linear response
transport regime.

As a further test of our conjecture, the temperature dependence of the
MW-induced $\Delta G$ is shown to be in qualitative agreement with models for
field assisted hopping. On the other hand, a quantitative analysis of these
non-Ohmic effects exposes a discrepancy between these theories and
experiments. This discrepancy, well documented in experiments on hopping
systems, manifests itself as a much higher sensitivity to fields than
suggested by equation 1. It is also pointed out that the exponential
dependence on the field, anticipated by theory, might hold only over a very
small range of fields.

\section{Experimental}

Several materials were used to prepare samples for measurements in this study.
These were thin films of In$_{\text{2}}$O$_{\text{3-x}}$ and In$_{\text{x}}$O,
and granular-aluminum. Samples from these materials were prepared by e-gun
evaporation of In$_{\text{2}}$O$_{\text{3}}$ or aluminum pellets unto room
temperature glass substrates in partial oxygen pressures of (4-6)$\cdot
$10$^{-5}$~mbar and rates of (0.3-{1)~\AA }/s. The In$_{\text{2}}%
$O$_{\text{3-x}}$ samples were obtained from the as-deposited amorphous films
by crystallization at 525~K. We also used for comparison samples of GaAs and
beryllium that were highly resistive and exhibited hopping conductance at
liquid helium temperatures.

Conductivity of most of the samples was measured using a two terminal ac
technique employing a 1211-ITHACO current preamplifier and a PAR-124A lock-in
amplifier. The GaAs specimen however, was configured for a Van-der-Pauw,
4-terminal ac measurement. In either case, the ac voltage bias was kept small
enough to minimize deviations from Ohmic conditions (as will be shown below,
some deviation may be observed even at the smallest fields used).

The high-power synthesizer HP8360B, was used with power up to 25~dBm
($\approx$316~mW) for the MW excitation. The frequency range used in this work
was limited to 2-6~GHz. The output of the synthesizer was fed to the sample
chamber via a coaxial cable ending with a short antenna. An important issue in
the study is the functional form of the sample response to the power of the
MW. It was therefore desired to ascertain the linearity of the synthesizer
setting versus its output as well as the integrity of the transmission line
itself. This was done by measuring the generated power near the instrument
output and near the sample stage. The results shown in figure 1 seem to rule
out the possibility that the sub-linearity is an instrumental artifact.
\begin{figure}
[ptb]
\begin{center}
\includegraphics[
trim=0.168903in 1.116324in 0.143881in 1.021214in,
height=2.5088in,
width=3.1739in
]%
{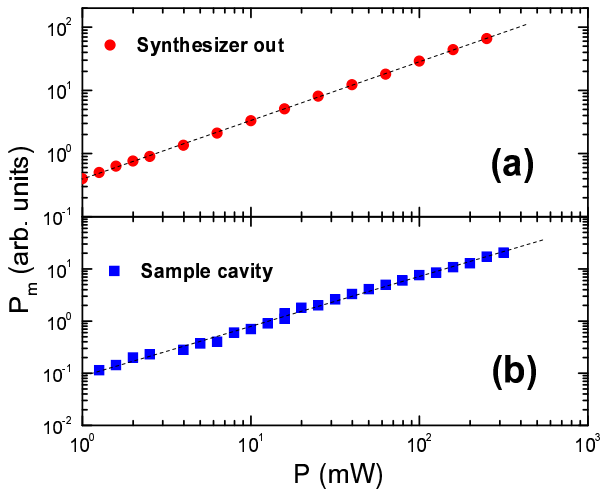}%
\caption{The power P$_{m}$ recorded by the power-meter as function of the
output power setting on the synthesizer P. The latter is used throughout the
paper to characterize MW power. Measurements shown are at a constant frequency
of 2.45~GHz; (a) measured about 2~cm from the synthesizer output socket. (b)
measured 1 cm from the short antenna at the sample-end of the transmission
line.}%
\end{center}
\end{figure}

Some auxiliary measurements described below employed the Tabor WS8101, a
100~MHz generator as the radio-frequency (RF) source. The output of this
generator was inserted into the sample cavity through the same transmission
line as the MW source. The linear relation between the setting of the
generators and the induced RF voltages across the sample was ascertained by
monitoring the pick-up waveforms on an oscilloscope. The data shown in this
paper are plotted with respect to the output settings of the respective RF or
MW source.

In both the MW and RF experiments an initial frequency scan was made to locate
a range where the response is conveniently large. Naturally that was usually
one of the resonances of the chaotic cavity in which the sample was mounted.
As will be shown below, the positions of these resonances was stable over many
hours, and reproducible results could be obtained with a barely noticeable
drift in frequency (%
$<$%
0.1\%/day).

Unless otherwise mentioned, measurements reported here were performed at
$T\approx$4.1K with the sample immersed in liquid helium. Complementary
details of sample preparation, characterization, and measurements techniques
are given elsewhere \cite{9}.

\section{Results and discussion}

The most characteristic feature of the MW-enhancement phenomenon is the
functional form of the excess conductance $\Delta G$ versus the radiation
power $P$. In more than 150 samples studied in our laboratory, $\Delta G$
increased with $P$ less than linearly. Results of $\Delta G(P)$ scans,
illustrating the sub-linear form of the response, are shown in figure 2. The
sub-linearity in these results are typical; cases where the sub-linearity was
less prominent than those in figure 2 were encountered usually only in samples
where the maximum effect was smaller than 2~\%.%
\begin{figure}
[ptb]
\begin{center}
\includegraphics[
trim=0.254919in 0.254699in 0.307310in 0.153142in,
height=3.4186in,
width=3.243in
]%
{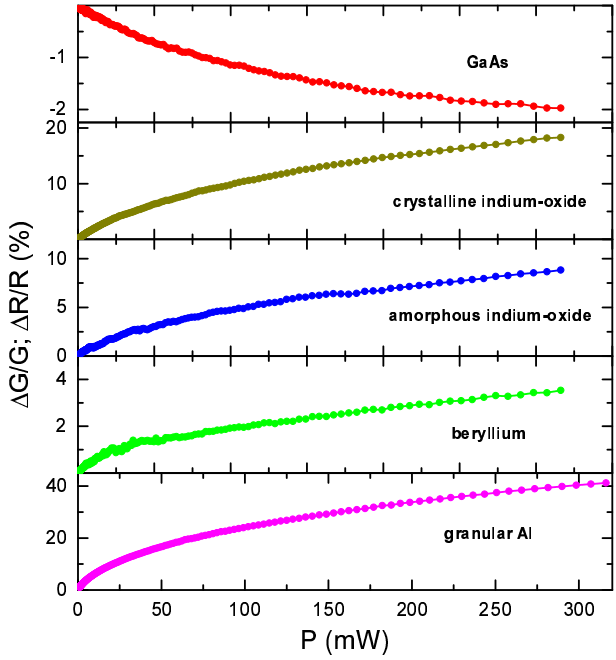}%
\caption{The fractional change of conductance (resistance in the case of the
GaAs sample, which was measured by a 4-terminal technique), as function of MW
power. All samples shown were measured at T=4.1K.}%
\end{center}
\end{figure}

While linearity at power levels smaller than measured here cannot be ruled
out, it is noted that there are instances where $\Delta G(P)$ is sub-linear at
a power such that $\Delta G/G\leq$10$^{-2}$. This was the first indication
that the origin of the effect is not consistent with `heating'; from the point
of view of power-balance, a spatially uniform increase of the electron
temperature will perforce give $\Delta G\propto P~$when both $\Delta T$ and
$\Delta G$ are small as is evidently the case for $\Delta G/G\leq$10$^{-2}$.
An even stronger evidence against `heating' was given in \cite{8} based on the
lack of change in the 'memory-dip' shape upon exposure to MW radiation. The
memory-dip is an identifying feature of intrinsic electron-glasses. It is a
cusp-like modulation observable in conductance versus gate voltage sweeps
\cite{10}. It has a shape that is highly sensitive to the electron temperature
and thus may be used a thermometer (see \cite{10} for details). The underlying
mechanism for the MW-enhanced conductivity was not identified in \cite{10}.

As more data became available, a statistical correlation emerged between the
sensitivity of $G$ to the voltage employed in the conductance measurement, and
the magnitude of $\Delta G$ produced by a given power of MW; samples that
showed Ohmic behavior up to relatively high voltages, exhibited small
MW-induced $\Delta G$, and \textit{vice-versa}.

The connection between the two effects became clear once their functional form
was compared. To affect this comparison the $\Delta G(P)$ is converted to
$\Delta G(b\cdot P^{1/2})$ where $b$ is a constant chosen to make an optimal
fit to $\Delta G(F)$ obtained independently from the current-voltage
characteristics of the sample. The similarity between the two functions, was
observed in all the tested samples, examples are shown in figure 3 and 4 below.

The following heuristic picture for the MW enhancement mechanism then suggests
itself: The MW radiation induces field across the sample \cite{11} (picked-up
by the wires connected to the sample for the two-terminal conductance
measurement). Once the potential difference associated with this field is
greater than the voltage employed for measuring $G$, the conductance of the
sample is modified in a similar vein as in measuring $G(F)$ directly (by dc or
low-frequency ac technique). In other words, the conductance monitored by a
low-bias, low-frequency technique takes advantage of the improved current-path
induced by the high-bias associated with the MW field. The reason for the
sub-linearity of $\Delta G(P)$ is then just the less-than-quadratic dependence
of the excess conductance on an applied field over the range relevant for the
measured $\Delta G$.%
\begin{figure}
[ptb]
\begin{center}
\includegraphics[
trim=0.162008in 1.098654in 0.294193in -0.046918in,
height=2.8686in,
width=3.2223in
]%
{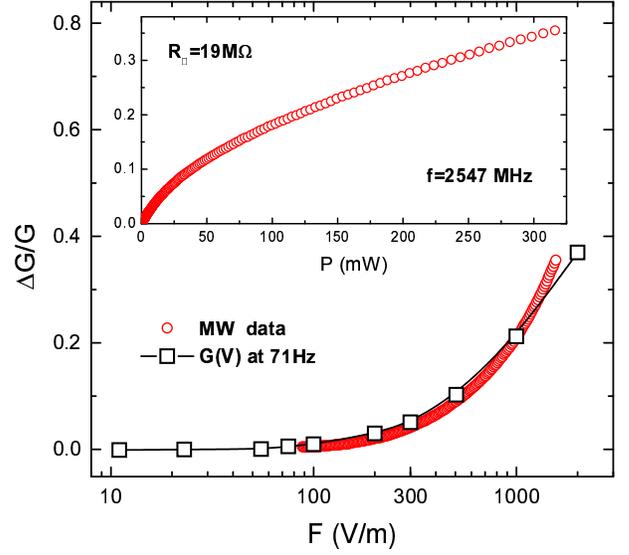}%
\caption{The fractional conductance change of a 210~\AA ~thick In$_{\text{x}}%
$O sample, 1x1mm$^{2}$ as function of the applied field. This is compared with
$\Delta G(P^{1/2})/G$ by adjusting its x-axis (see text). The inset shows
$\Delta G(P)/G$ of this sample.}%
\end{center}
\end{figure}
%

\begin{figure}
[ptb]
\begin{center}
\includegraphics[
trim=0.139189in 0.859207in 0.173595in 0.145082in,
height=2.9914in,
width=3.1808in
]%
{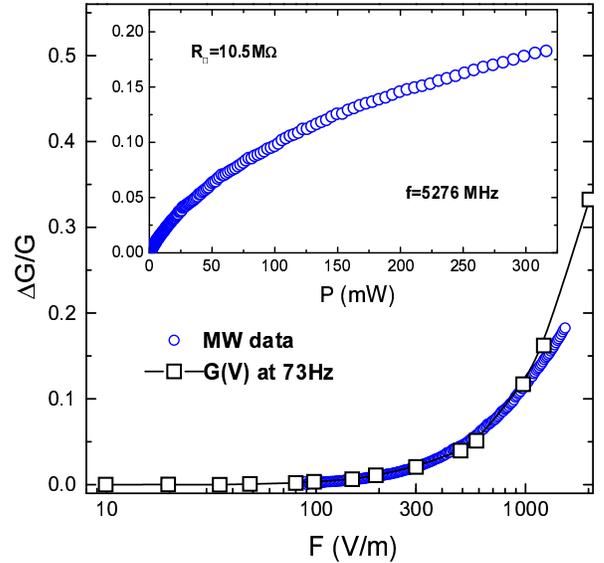}%
\caption{Same as in figure 3 for a 50~\AA ~thick In$_{\text{2}}$%
O$_{\text{3-x}}$ sample.}%
\end{center}
\end{figure}

$\Delta G(F)$ of a given sample is similar but it does not perfectly match the
respective $\Delta G(b\cdot P^{1/2})$ (see, for example, figures 3 and 4).
There were always some degree of imprecise registry between these functions in
all the samples we tested. Such deviations ought perhaps to have been
expected. The assumption, implicit to our matching procedure, that it is only
the \textit{amplitude} of the field that matters, is inaccurate; the
Miller-Abrahams \cite{12} resistors comprising the hopping system are not
purely resistive, and therefore, the local potential drops across the sample
may be different at different frequencies.

To elucidate this point, we measured $\Delta G[F(\omega)]$ at different
frequencies on several samples. $F$($\omega$) was the modulating field applied
at frequency $\omega$ while the conductance was monitored at some
low-frequency (typically, 20-75~Hz) as in the MW experiments. Here however, we
employed frequencies in the 10$^{6}$-10$^{8}$~Hz range so the voltages induced
across the sample could be readily displayed and measured by an oscilloscope.
Figure 5 shows results of such an experiment. The 3.3~MHz and 15.52~MHz curves
in the figure were adjusted to coincide with the 23~Hz plot at $\Delta
G/G$=0.7 by re-scaling their voltage axis by a constant, just as in the MW
case. This illustrates that the $\Delta G(F)$ is also a function of the field
frequency. Interestingly, above a certain frequency the constant used for
re-scaling the $\Delta G[F(\omega)]$ curves was somewhat smaller as $\omega$
became higher. Apparently, to achieve the same $\Delta G/G$, larger amplitude
is needed at higher $\omega$. This issue is currently under investigation.%

\begin{figure}
[ptb]
\begin{center}
\includegraphics[
trim=0.139971in 0.993810in 0.200964in 0.157978in,
height=3.0113in,
width=3.2578in
]%
{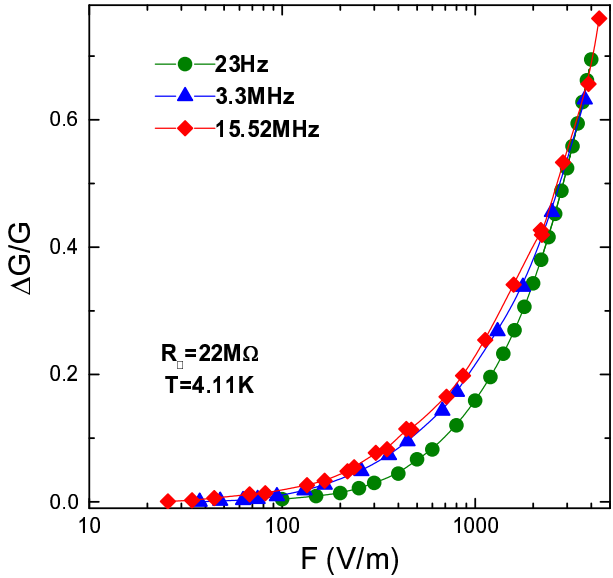}%
\caption{The fractional change of conductance as function of field (measured
at 23~Hz by a two-terminal lock-in technique). The two other curves were
measured on the same sample, by exposing it to radiation at the indicated
frequency. The field-axis of the high frequency curves was adjusted to match
the 23~Hz curve at $\Delta G/G$=0.7. The sample is In$_{\text{2}}%
$O$_{\text{3-x}}$ with thickness 43~\AA .}%
\end{center}
\end{figure}

The conjecture that the MW-enhanced conductance is a field-induced non-Ohmic
effect is also supported by the temperature dependence of the excess
conductance (at constant MW power). In these experiments, $\Delta G(\omega)$
was measured over a range of frequencies at each temperature. This was done to
cater for drifts in the cavity characteristics with the concomitant shift in
conductance-response peaks (see figure 6). The actual drift observed in the
figure is remarkably small; note that the series of plots were collected over
3 days because at each temperature the sample was allowed to equilibrate for
several hours (the sample is an electron-glass with slow relaxation times
\cite{13}).

The inset to the figure shows $\Delta G(T)$ due to the MW radiation normalized
by $G(T)$ at zero field. These experimental points fit reasonably well a
power-law which may actually be in qualitative agreement with theory:
According to Pollak and Riess \cite{4}, and Shklovskii \cite{3}, at fields
just above the "Ohmic regime" the conductance versus field is given by:%
\begin{equation}
G(F,T)=G(T)\cdot\exp\left[  \frac{eFL(T)}{k_{_{B}}T}\right]  \label{2}%
\end{equation}

and therefore:%

\begin{equation}
\frac{\Delta G(F)}{G}=\exp\left[  \frac{eFL(T)}{k_{_{B}}T}\right]  -1
\label{2a}%
\end{equation}

For small $\Delta G/G$ this yields:%
\begin{equation}
\frac{\Delta G}{G}\propto\frac{eFL(T)}{k_{_{B}}T} \label{3}%
\end{equation}

Pollak and Riess associate $L(T)$ with the hopping length \cite{4} while
Shklovskii associates it with the percolation radius \cite{3}. In either case
one expects a power law dependence as indeed observed (see, inset to figure
6). To push the analysis a little further, note that the resistance versus
temperature of this sample exhibits Mott's variable-range-hopping \cite{14} of
a two-dimensional system (figure 7). In this case the hopping length $r$
scales with temperature as $r(T)\propto T^{-1/3}$ and the percolation-radius
$L_{C}$\ scales like $L_{C}(T)\propto T^{-2/3}$ giving $\Delta G/G\propto
T^{-1.33}$ and $\Delta G/G\propto T^{-1.67}$ respectively. The fit to the data
in the inset to figure 6 yields $\Delta G/G\propto T^{-1.75}$ which may
suggest that $L_{C}$\ is the relevant length.%
\begin{figure}
[ptb]
\begin{center}
\includegraphics[
trim=0.280724in 0.983332in 0.329205in 0.145888in,
height=3.1695in,
width=3.2958in
]%
{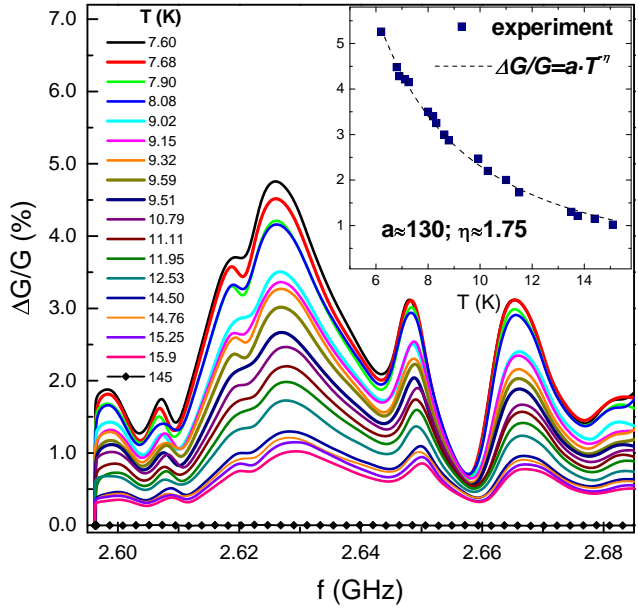}%
\caption{Isotherm plots for the fractional change of conductance versus the MW
frequency. Data are taken under a constant power of 25~dBm. The sample is
50~\AA  ~In$_{\text{2}}$O$_{\text{3-x}}$ film. Note the shift of the $\Delta
G(f)$ peak positions with temperature. The inset shows the temperature
dependence of $\Delta G(f)/G$ averaged over the range: $f$=2628$\pm$20~MHz (to
take care of the shift with temperature of the resonance peaks). The dashed
line is a power-law fit to the data (c.f., equation 3).}%
\end{center}
\end{figure}

The consistency with theory, as to temperature dependence, encouraged us to
test the agreement of the theory \cite{3} with the experiment in a more
detailed way. Using the relevant parameters from the $R(T)$ data of the sample
(figure 7), we estimate the percolation-radius, $L^{\ast}\approxeq\xi
\cdot(T_{0}/T)^{2/3}\approx$5000~\AA ~at 4K ($\xi$ is the
localization-length). This value of $L_{C}$\ is used in equation 2a to plot
$\Delta G(F)/G$ and the result is compared with the experimentally measured
values in figure 8.

This comparison brings to light several problems that need be addressed. The
theory \textit{may} fit the experiment for very small $\Delta G(F)/G$,
however, it systematically deviates from the experimental curve when $\Delta
G(F)/G\gtrsim$6~\%. This discrepancy is not due to the uncertainty in the
value of $L_{C}$: The experimental curve is simply \textit{not} exponential
over the range where the effect is more than few percents, (which makes the
use of this function questionable). Nor is the range of the studied fields out
of the `intermediate' regime. The field where the problem appears is still
\textit{much} smaller than the high-field regime, where the conductance
becomes temperature independent \cite{5,15}. On the other hand, if we accept
that the fit from $F\approx$1~V/m to $F\approx$80~V/m (see inset to figure 8)
is in agreement with theory, we have a problem accounting for the condition
for the onset of the exponential dependence: This is expected to be given by
equation 1, and inserting $F\approx$1~V/m gives $L\approx$400$~\mu$m. It is
difficult to assign a physical meaning to such a length-scale in a hopping
system at $\approx$4K.%

\begin{figure}
[ptb]
\begin{center}
\includegraphics[
trim=0.242609in 2.237969in 0.336106in 0.905510in,
height=2.156in,
width=3.4307in
]%
{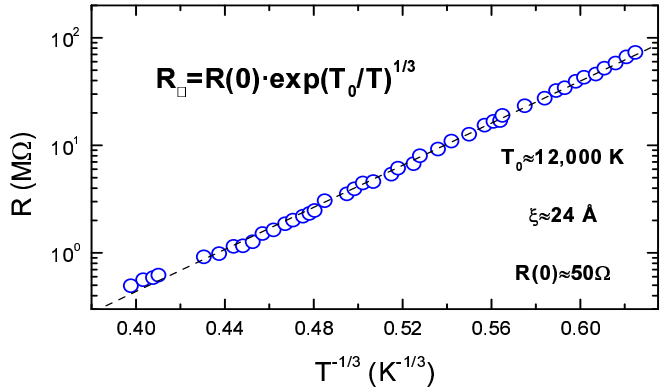}%
\caption{Resistance as function of temeparture for the same In$_{\text{2}}%
$O$_{\text{3-x}}$ sample shown in figure 6.}%
\end{center}
\end{figure}

This problem is commonly encountered in hopping conductivity. Usually, the
puzzle is presented in terms of length-scale (being the one parameter in
equation 1 that is not "measured"). Values of this length that are larger by
1-3 orders of magnitude than reasonably expected are reported, or could be
inferred, from data in the literature \cite{16,17,18}. These include results
for Ge samples doped by nuclear-transmutation \cite{17}, presumed to be
relatively free of technical inhomogeneities. Remarkably, the field where
non-Ohmicity was already evident in these experiments was smaller by at least
\cite{17} a factor of 25 than the theoretical value. These authors \cite{17}
find agreement with the Shklovskii model \cite{3} in terms of the temperature
dependence of $L(T)$, however, their data fit equation 2 only over a very
small range ($\leq$6~\%).%

\begin{figure}
[ptb]
\begin{center}
\includegraphics[
trim=0.172486in 2.068284in 0.359481in 0.920367in,
height=2.1378in,
width=3.3217in
]%
{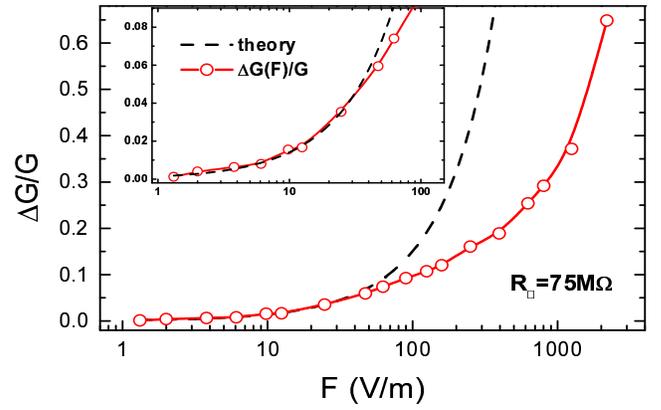}%
\caption{The fractional change of the conductance as function of applied field
for the In$_{\text{2}}$O$_{\text{3-x}}$ sample shown in figure 6. The inset is
a zoomed-in view of the small field region illustrating that non-Ohmicity is
evident at very small fields. The "self-similar" impression one gets by
comparing the zoomed-in view with the data in the main figure is a graphical
testimony to that $\Delta G(F)$ is not an exponential function.}%
\end{center}
\end{figure}

It has been remarked by several authors \cite{16,15} that the origin of the
discrepancy between theory and experiment as to the onset-field for
non-Ohmicity may be a result of non-uniform field distribution. This might
make the field effectively larger than the field one uses in equation 1 based
on the measured potential difference across the sample divided by its length.
Spatial inhomogeneity is inherent to Anderson insulators, and at least
partially, this is actually embodied in the theory in that the system is
self-averaging only on scales larger than $L_{C}$, which may be quite large at
low temperatures. The observation that there has to be a larger length scale
to account for the experimental results on different systems suggests that
there is an inherent reason not taken into account in the standard hopping
models (in addition to imperfections due to technological reasons). There
seems to be evidence, based on studies of the onset of non-Ohmic behavior as
function of sample-size \cite{19}, that long-range potential-fluctuations may
play a role in this phenomenon.

Note, incidentally, that the sub-linearity of $\Delta G(P)$ at small fields
follow naturally from theoretical models that predict $G(F)\propto\exp
[\frac{eFL}{k_{B}T}]$ \cite{1,3,4}. The Apsley and Hughes model \cite{2} that
expects $G(F)\propto\exp[(\frac{eFL}{k_{B}T})^{2}]$ yields a linear dependence
at small $P,~$however, it does not fit the measured $G(F)$ any better than the
other models. It is unfortunate that neither model may be trusted to fit
experiments on real systems except over a limited range of fields. Given this
situation, comparing the MW-enhanced conductance with the experimentally
measured $G(F)$ is the only option one has for a meaningful test.

The similarity between the field assisted and the MW-enhanced conductance and
the qualitative agreement with theory make a strong case for the mechanism we
propose. This is a plausible scenario for the MW frequency regime employed in
this study as the associated wavelength is larger than the sample size.

\section{Acknowledgments}

This work greatly benefited from the insightful remarks by M. Pollak and from
the many helpful discussions with B. Shapiro during the Electron-Glasses
program held at Santa-Barbara. The author expresses gratitude for the
hospitality and support of the KAVLI institute that hosted this program.
Illuminating discussions with A. Frydman and A. Goldman at the later stage of
this work are gratefully acknowledged. Thanks are also due to X. M. Xiong, and
P. W. Adams for the Be samples, and to N.V. Agrinskaya for the GaAs sample.
This research was supported by a grant administered by the US Israel
Binational Science Foundation and by the Israeli Foundation for Sciences and Humanities.


\begin{thebibliography}{99}
\bibitem {11}The MW-induced voltage across the sample (with $\omega
>$10$^{\text{10}}$~s$^{\text{-1}}$) could not have been measured in this work.
\end{thebibliography}
\end{document}